\documentclass{llncs}

\usepackage{verbatim}
\usepackage[utf8x]{inputenc}
\usepackage{graphicx}
\usepackage{amsfonts}
\usepackage{amssymb}
\usepackage{amsmath}
\usepackage{latexsym}
\usepackage{amsmath}
\usepackage{amsmath}
\usepackage{amssymb}
\usepackage{stmaryrd}
\usepackage{graphicx}
\usepackage{algorithm}
\usepackage{algorithmic}
\usepackage{subfigure}
\usepackage{amsfonts,amssymb,latexsym}
\usepackage{yfonts}
\usepackage{color}
\usepackage[all]{xy}
\usepackage{wrapfig}
\usepackage{cases}
\usepackage{gastex,amsfonts,amssymb,latexsym}
\usepackage{times}
\usepackage{gastex}
\usepackage{amsmath}
\usepackage{amssymb}
\usepackage{stmaryrd}

\usepackage{algorithm}
\usepackage{algorithmic}
\usepackage{subfigure}
\usepackage{gastex,amsfonts,amssymb,latexsym}
\usepackage{yfonts}
\usepackage{color}
\usepackage[all]{xy}
\usepackage{arrows}
\usepackage{wrapfig}
\usepackage{tabularx,colortbl}

\renewcommand{\emptyset}{\varnothing}

\newcommand{\mc}[1]{\mathcal{#1}}

\newcommand{\SP}{\textsc{SP}}

\newcommand{\mv}[1]{\singlearrow{#1}}

\renewcommand{\P}{{\bf P}}

\newcommand{\AP}{\textsc{AP}}

\newcommand{\Paths}{{\it Paths}}

\renewcommand{\tt}{\textrm{tt}}

\newcommand{\Reals}{\mathbb{R}}

\newcommand{\real}{\Reals}

\def\topbotatom#1{\hbox{\hbox to 0pt{$#1\bot$\hss}$#1\top$}}

\makeatletter
\newcommand{\be}{%
\begin{group}
\eqnarray%
\@ifstar{\nonumber}{}%
}

\makeatother

\usepackage{amsfonts}
\usepackage{amsmath}
\usepackage{amstext}
\usepackage{amssymb}
\usepackage{times}

\usepackage{graphicx}
\graphicspath{{figures/}}
\usepackage{subfigure}

\usepackage{algorithm}
\usepackage{algorithmic}
\usepackage{wrapfig}
\usepackage[all]{xy}
\usepackage{hyperref}
\usepackage{breakurl}

\usepackage{gastex}

\begin{document}
\title{A Two Step Perspective for Kripke Structure Reduction}



\author{Arpit Sharma}
\institute{Software Modeling and Verification Group, RWTH Aachen University, Germany \\
arpit.sharma@cs.rwth-aachen.de}
\maketitle
\begin{abstract}
This paper presents a novel theoretical framework for the state space reduction of Kripke structures. We define two equivalence relations, Kripke minimization equivalence (KME) and weak Kripke minimization equivalence (WKME). We define the quotient system under these relations and show that these relations are strictly coarser than strong (bi)simulation and divergence-sensitive stutter (bi)simulation, respectively. We prove that the quotient system obtained under KME and WKME preserves linear-time and stutter-insensitive linear-time properties. Finally, we show that KME is compositional w.r.t. synchronous parallel composition.      
\end{abstract}
\section{Introduction}
Model checking Kripke structures (KSs) \cite{KB08} suffers from the well-known \emph{state-space explosion} problem where the number of states grows exponentially in the number of parallel components. Abstraction techniques based on equivalence relations reduce the state space of KSs, by aggregating equivalent states into a single state. The reduced state space obtained under an equivalence relation, called a quotient, can then be used for analysis provided it preserves a rich class of properties of interest. For KSs, one usually distinguishes between linear-time and branching-time equivalence relations \cite{RJG01}. The standard example of a linear-time equivalence is trace equivalence \cite{Hoare78,Rem87,JLA85}. Informally, two states are trace equivalent if the possible sequences of words starting from these states are the same. Several extensions of trace equivalence have been proposed, e.g., failure semantics and readiness semantics \cite{CSAR84,CAR85,RDNMH84,Nicola87,CH86,APnueli85,BBWK87,LPomello85,VeglioniN98}. In the weak setting, stutter trace equivalence has been proposed where a pair of sequences are considered to be equivalent if they differ in at most the number of times a set of propositions may adjacently repeat \cite{LLamport83}. Checking trace equivalence is PSPACE-complete. In branching-time semantics, various relations on KSs have been defined such as strong and stutter variants of bisimulation and simulation pre-orders \cite{MilnerR80,DPark81,Milner71,JFGrooteV90,EBrowneCG88,WPWGlabbeekW96}. Strong bisimulation and divergence-sensitive stutter bisimulation coincide with Computation Tree Logic $(CTL^{*})$ and $CTL^{*}/_{\bigcirc}$, respectively \cite{EBrowneCG88,FWVNicolaV90a}. Strong simulation agrees with a “preorder” on the universal (or existential) fragment of $CTL$ \cite{ODClarkeGL94}. Several papers report data showing that bisimulation minimization can substantially reduce the state-space of models to be verified \cite{AVSRKB94,KFMYV98}. The use of simulation relations for abstraction has been studied in, e.g., \cite{ODClarkeGL94,PCousotC02,LoiseauxGSBB95}. Unfortunately, (stutter) (bi)simulation is too fine, and it is often desirable to obtain a quotient system smaller than (stutter) (bi)simulation such that properties of interest are still preserved. This is particularly important if the properties to be verified belong to the class of (stutter-insensitive) linear-time properties, e.g. safety properties, liveness properties and in general (stutter-insensitive) $\omega$-regular properties. These properties can be expressed using temporal logics such as Linear Temporal Logic $(LTL)$ \cite{KB08}, Property Specification Language $(PSL)$ \cite{IEEE05} and  semi-extended $PSL$ $(siPSL)$ \cite{DaxKL09}.   

In this paper our focus is on Kripke minimization equivalence (KME) that allows for a more aggressive state space reduction than strong (bi)simulation. In the weak setting we define weak Kripke minimization equivalence (WKME) such that state space reduction under WKME can potentially be much larger than for divergence-sensitive stutter (bi)simulation. Whereas bisimulation compares states on the basis of their direct successors, KME considers a \emph{two-step} perspective. Two states $s$ and $s'$ are KME equivalent if for each pair of their direct predecessors it is possible to directly move to any equivalence class via the equivalence class $[s]=[s']$. The main principle is captured in Fig. 1 where only those states can be merged into equivalence class $C$ for whom $s_{p}$ and $s'_{p}$ can reach equivalence classes $D$ and $E$ via $C$ and this should hold for each pair of predecessors of $C$. Intuitively, each predecessor of $C$ should reach the same set of equivalence classes in two steps via $C$. In Fig. 1 it may be possible that some of these predecessors have only one successor in $C$ while others have multiple successors in $C$. For WKME, we abstract from stutter steps and thus each predecessor of $C$ should reach the same set of equivalence classes in two or more steps such that all extra steps are taken within $C$.    
\begin{figure}
 \centering
\scalebox{0.6}{
\Large\begin{picture}(74,58)(0,-58)
\node[Nfill=y,fillcolor=Black,ExtNL=y,NLangle=0.0,Nw=3.82,Nh=3.28,Nmr=1.64](n0)(28.0,-8.0){$s_{p}$}
\node[Nfill=y,fillcolor=Black,ExtNL=y,NLangle=0.0,Nw=3.82,Nh=3.28,Nmr=1.64](n7)(44.0,-8.0){$s'_{p}$}
\node[NLangle=0.0,Nw=25.13,Nh=9.83,Nmr=2.46](n10)(36.0,-28.0){$C$}
\node[NLangle=0.0,Nw=25.13,Nh=9.83,Nmr=2.46](n14)(16.0,-48.0){$D$}
\drawbpedge(n10,-161,1.73,n14,-180,3.0){}
\drawedge(n0,n10){}
\drawedge(n7,n10){}
\node[NLangle=0.0,Nw=25.13,Nh=9.83,Nmr=2.46](n23)(56.0,-48.0){$E$}
\drawedge(n10,n23){}
\end{picture}}
\caption{Kripke minimization equivalence}
\end{figure}
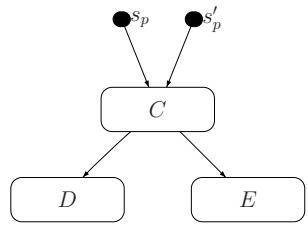 
\paragraph*{Contributions.} The main contributions of this paper are as follows:
\begin{itemize}
 \item We provide a structural definition of KME on KSs, define the quotient under KME and show that KME is strictly coarser than strong (bi)simulation.
 \item We show that linear-time (LT) properties defined over infinite words are preserved under KME quotienting. 
 \item In the weak setting, we provide a structural definition of WKME on KSs, define the quotient under WKME and show that WKME is strictly coarser than divergence-sensitive stutter (bi)simulation.
 \item Next, we prove that stutter-insensitive LT properties defined over infinite words are preserved under WKME quotienting. 
 \item Finally, we show that KME is compositional w.r.t. synchronous parallel compositon (SCCS-like parallel composition \cite{CCMilner83}).    
\end{itemize}
The theory presented in this paper forms the basis for developing an efficient algorithm that can obtain quotient systems that are smaller than (stutter) (bi)simulation. This is particularly helpful in situations where several components have to be combined using synchronous parallel composition \cite{CCMilner83}, as KME based reduction can be applied at each step of the iterative composition. Both KME and WKME defined in this paper can be seen as state space reduction techniques induced by trace equivalence and stutter trace equivalence, respectively.                                      
\paragraph*{Related work.} In the stochastic context, T-Lumpability has been defined over sequential Markovian process calculus (SMPC) \cite{BM08}. T-Lumpability is defined using four process-algebraic axioms, and allows for a more aggressive state space aggregation than ordinary lumpability. In \cite{SK11} a novel structural definition of  weighted lumpability (WL) has been provided on continuous-time Markov chains (CTMCs) that coincides with T-Lumpability. For WL it has been proved that probability of properties specified using deterministic timed automaton and metric temporal logic are preserved under WL quotienting. Recently, the notion of WL has been extended to discrete-time Markov chains (DTMCs) and the preservation result for probability of $\omega$-regular properties has been established \cite{Sharma12}. Our definition of equivalence for strong case, i.e., KME here builds on that investigated in \cite{SK11} for CTMCs.                       
\paragraph*{Organisation of the paper.}
Section 2 briefly recalls the basic concepts of KSs. 
Section 3 defines Kripke minimization equivalence and discusses the preservation of LT properties under KME quotienting. 
Sections 4 defines weak Kripke minimization equivalence and discusses the preservation of stutter-insensitive LT properties under WKME quotienting. In section 5, we prove that WPE is compositional w.r.t. synchronous parallel composition. Finally, section 6 concludes the paper.
\section{Preliminaries}
This section recalls the basic concepts of Kripke structures with a finite state space.  
\begin{definition}[KS]
A Kripke structure (KS) is a tuple $\mathcal{K}=(S,\rightarrow,AP,L,s_{0})$ where:
\begin{itemize}
\item $S$ is a non-empty finite set of states,
\item $\rightarrow \subseteq S\times S$, is a transition relation s.t. $\forall s \in S \exists s'\in S$ with $(s,s')\in \rightarrow$,
\item $AP$ is a finite set of atomic propositions,
\item $L:S\rightarrow2^{AP}$ is a labeling function,
\item $s_{0}\in S$ is the initial state.
\end{itemize} 
\end{definition}
For simplicity, we write $s\mv{}s'$ instead of $(s,s')\in$ $\rightarrow$. Let $s\in S$ and $C\subseteq S$, then $Post(s,C)=\{s'\in C\mid s\mv{}s'\}$. Let $Post(s) = \{ s' \in S \mid s\mv{}s' \}$. For $C\subseteq S$, let $pred(C)=\{s' \mid \exists s\in C.s'\mv{} s\}$. 
\begin{definition}[KS paths]
Let $\mathcal{K}=(S,\rightarrow,AP,L,s_{0})$ be a KS. An infinite \emph{path} $\pi$ in $\mathcal{K}$ is an infinite state sequence, i.e., $s_0\mv{}s_1\mv{}s_2\ldots \in S^{\omega}$ with $s_{i}\in S$.
\end{definition}
Note that, since we do not allow KS $\mathcal{K}$ to have terminal states, i.e., which do not have any outgoing transitions, we only consider infinite paths (starting from the initial state).  
Let $\Paths^{\mathcal{K}}(s_{0})$ denote the set of all infinite paths in $\mc{K}$ that start in $s_{0}$. For infinite path $\pi$ and any $i\in \mathbb{N}$, let $\pi[i]=s_{i}$, the $(i+1)$-st state of $\pi$. Let $\pi[i...]$ denote the suffix of path $\pi$ starting in the $(i+1)$-st state.
\begin{definition}[KS traces]
Let $\mathcal{K}=(S,\rightarrow,AP,L,s_{0})$ be a KS. The \emph{trace} of an infinite path $\pi=s_0\mv{}s_1\mv{}s_2\ldots \in S^{\omega}$ is $trace(\pi)=L(s_{0})L(s_{1})L(s_{2})\ldots \in (2^{AP})^{\omega}$. 
\end{definition}
Intuitively a trace of an infinite path is the infinite sequence of sets of atomic propositions that are valid in the states of the path, i.e. an infinite word over the alphabet $2^{AP}$. Let $Traces^{\mathcal{K}}(s_{0})$ denote the set of all infinite traces in $\mc{K}$ that start in $s_{0}$.
\begin{definition}[Trace-equivalent paths]
Let $\mathcal{K}=(S,\rightarrow,AP,L,s_{0})$ be a KS and $\pi_{i}\in \Paths^{\mathcal{K}}(s_{0}),$ $i=1,2$. $\pi_{1}$ and $\pi_{2}$ are \emph{trace-equivalent}, denoted by $\pi_{1}\triangle \pi_{2}$, if $L(\pi_{1}[i])=L(\pi_{2}[i])$ for all $i\ge 0$. 
\end{definition}
\begin{definition}[Stutter step]
Transition $s \mv{} s'$ in Kripke structure $\mathcal{K}=(S,\rightarrow,AP,L,s_{0})$ is a \emph{stutter step} if $L(s)=L(s')$.  
\end{definition}
The notion of stuttering is lifted to paths as follows.
\begin{definition}[Stutter-equivalent paths]
Let $\mathcal{K}=(S,\rightarrow,AP,L,s_{0})$ be a KS and $\pi_{i}\in \Paths^{\mathcal{K}}(s_{0}),$ $i=1,2$. $\pi_{1}$ and $\pi_{2}$ are stutter-equivalent, denoted by $\pi_{1}\triangleq \pi_{2}$, if there exists an infinite sequence $A_{0}A_{1}A_{2}\ldots$ with $A_{i}\subseteq AP$ and natural numbers $n_{0},n_{1},n_{2},\ldots,$ $m_{0},m_{1},m_{2},\ldots \ge 1$ s.t. 
\begin{equation*}
 trace(\pi_{1})=\underbrace{A_{0}\ldots A_{0}}_{n_{0}-times}\underbrace{A_{1}\ldots A_{1}}_{n_{1}-times}\underbrace{A_{2}\ldots A_{2}}_{n_{2}-times}\ldots
\end{equation*}
 \begin{equation*}
 trace(\pi_{2})=\underbrace{A_{0}\ldots A_{0}}_{m_{0}-times}\underbrace{A_{1}\ldots A_{1}}_{m_{1}-times}\underbrace{A_{2}\ldots A_{2}}_{m_{2}-times}\ldots
\end{equation*}  
where $\underbrace{A_{0}\ldots A_{0}}_{n_{0}-times}$ denotes for all $i=0\ldots n_{0}-1$, $L(\pi_{1}[i])=A_{0}$.       
\end{definition}
Note that $\underbrace{A_{0}\ldots A_{0}}_{n_{0}-times}$ only refers to the first block, for other blocks it is defined in an analogous manner. Accordingly, stutter-equivalence for any two infinite traces $\rho_{1},\rho_{2}\in (2^{AP})^{\omega}$ (denoted by $\rho_{1}\triangleq \rho_{2}$) can be defined. 
\begin{example}
Consider the KS $\mathcal{K}$ in Fig. 2 (left), where $S=\{s_{0},s_{1},s_{2},s_{3},s_{4},s_{5},s_{6},s_{7}\}$, $AP=\{a,b\}$ and $s_{0}$ is the initial state. An example finite path $\pi$ is $s_{0}\mv{}s_{1}\mv{}s_{4}\mv{}s_{6}$. Here $\pi[3]=s_{6}$. The trace for $\pi$ is given by $trace(\pi)=\{a\}\emptyset\{a\}\{b\}$. 
\end{example}

\paragraph{Assumptions.}
Throughout this paper we assume that every state of KS $\mathcal{K}$ has at least one predecessor, i.e., $pred(s) = \{ s' \in S \mid s'\mv{}s \} \ne \emptyset$ for any $s \in S$. 
This is not a restriction, as any KS $(S,\rightarrow,AP,L,s_{0})$ can be transformed into an equivalent KS $(S',\rightarrow^{'},AP',L',s'_{0})$ which fulfills this condition.
This is done by adding a new state $\hat{s}$ to $S$ equipped with a self-loop and which has a transition to each state in $S$ without predecessors.
To distinguish this state from the others we set $L'(\hat{s}) = \bot$ with $\bot \not\in \AP$.
(All other labels, states and transitions remain unaffected.) Let $s'_0 = s_0$.
It follows that all states in $S' = S \cup \{ \hat{s} \}$ have at least one predecessor.
Moreover, the reachable state space of both KSs coincides.
We also assume that the initial state $s_{0}$ of a KS is distinguished from all other states by a unique label, say $\$$.
This assumption implies that for any equivalence that groups equally labeled states, $\{ s_0 \}$ constitutes a separate equivalence class.
Both assumptions do not affect the basic properties of the KS such as linear or branching-time properties. 
For convenience, we neither show the state $\hat{s}$ nor the label $\$$ in figures.  
\section{Kripke Minimization Equivalence}
In this section, we present a technique for the state space minimization of a KS. We first define \emph{Kripke minimization equivalence} (KME) followed by the definition of quotient KS under KME. Next to that, the relationship between KME and strong (bi)simulation is explored.   
\begin{definition}[Predecessor based reachability]
For $s\in S$ and $C,D \subseteq S$, the function $Pbr:S\times2^{S}\times 2^{S}\rightarrow \{0,1\}$ is defined as:
\begin{equation*}
Pbr(s,C,D)=
\left\{\begin{array}{ll}
1 & \mbox{if } \exists s'\in Post(s,C) \mbox{ s.t. } \\ 
  & Post(s',D)\neq \emptyset\\ 
0 & \mbox{otherwise.} 
\end{array}\right.
\end{equation*} 
\end{definition}
\begin{definition}[KME]
Equivalence $\mathcal{R}$ on $S$ is a \emph{Kripke minimization equivalence} (KME) on $\mathcal{K}$ if we have: 
\begin{enumerate}
\item 
$\forall (s_{1},s_{2}) \in \mathcal{R}$ it holds: $L(s_{1}) = L(s_{2})$ and
\item $\forall C,D \in S/_{\mathcal{R}}$ and $\forall s',s''\in pred(C)$ it holds:
$Pbr(s',C,D) = Pbr(s'',C,D)$
\end{enumerate}
States $s_{1}$, $s_{2}$ are \emph{Kripke minimization equivalent}, denoted by $s_{1}\star s_{2}$, if $(s_{1},s_{2})\in \mathcal{R}$ for some KME $\mathcal{R}$.
\end{definition}
\begin{example}
Consider the KS $\mathcal{K}$ in Fig. 2 (left). Let $C=\{s_{3},s_{4},s_{5}\}$ and $D=\{s_{7}\}$. Then $Pbr(s_{1},C,D)=1$, since it is possible to move from $s_{1}$ to $s_{7}$ in two steps via $s_{3}$. Similarly $Pbr(s_{2},C,D)=1$. For KS $\mathcal{K}$, the equivalence relation induced by the partitioning $\{\{s_{0}\},\{s_{1}\},\{s_{2}\},\{s_{3},s_{4},s_{5}\},\{s_{6}\},\{s_{7}\}\}$ is a KME.   
\end{example}

\begin{definition}[Quotient Kripke structure]
For KME relation $\mathcal{R}$ on $\mathcal{K}$, the quotient Kripke structure $\mathcal{K/}_{\mathcal{R}}$ is defined by $\mathcal{K/}_{\mathcal{R}}=(S/_{\mathcal{R}},\rightarrow^{'},AP,L',s'_{0})$ where:
\begin{itemize}
\item 
$S/_{\mathcal{R}}$ is the set of all equivalence classes under $\mathcal{R}$,
\item $\rightarrow^{'}\subseteq S/_{\mathcal{R}}\times S/_{\mathcal{R}}$ is defined by: $C\mv{}^{'}D$ iff $Pbr(s',C,D)=1$ where $s'\in pred(C)$ and $C,D\in S/_{\mathcal{R}}$, 
\item $L'(C)=L(s)$, where $s\in C$ and
\item $s'_{0}=C$ where $s_{0} \in C$.
\end{itemize}
\end{definition} 
\begin{example}
The quotient KS for the Fig. 2 (left) under the KME relation with partition $\{\{s_{0}\},\{s_{1}\},\{s_{2}\},\{s_{3},s_{4},s_{5}\},\{s_{6}\},\{s_{7}\}\}$ is shown in Fig. 2 (right).
\end{example}
\begin{figure}
\centering
\scalebox{0.7}{
\Large \begin{picture}(99,79)(0,-79)
\node[NLangle=0.0](n0)(24.0,-8.0){$s_{0}$}
\nodelabel[ExtNL=y,NLangle=0.0](n0){$\{a\}$}
\node[NLangle=0.0](n1)(12.0,-20.0){$s_{1}$}
\nodelabel[ExtNL=y,NLangle=0.0](n1){$\{\}$}
\node[NLangle=0.0](n2)(36.0,-20.0){$s_{2}$}
\nodelabel[ExtNL=y,NLangle=0.0](n2){$\{b\}$}
\node[NLangle=0.0](n3)(4.0,-36.0){$s_{3}$}
\nodelabel[ExtNL=y,NLangle=0.0](n3){$\{a\}$}
\node[NLangle=0.0](n4)(24.0,-36.0){$s_{4}$}
\nodelabel[ExtNL=y,NLangle=0.0](n4){$\{a\}$}
\node[NLangle=0.0](n5)(44.0,-36.0){$s_{5}$}
\nodelabel[ExtNL=y,NLangle=0.0](n5){$\{a\}$}
\node[NLangle=0.0](n6)(16.0,-60.0){$s_{7}$}
\nodelabel[ExtNL=y,NLangle=0.0](n6){$\{\}$}
\node[NLangle=0.0](n7)(36.0,-60.0){$s_{6}$}
\nodelabel[ExtNL=y,NLangle=0.0](n7){$\{b\}$}
\drawedge(n0,n1){}
\drawedge(n0,n2){}
\drawedge(n1,n4){}
\drawedge(n2,n4){}
\drawedge(n2,n5){}
\drawedge(n1,n3){}
\drawedge(n4,n7){}
\drawedge(n3,n6){}
\drawedge(n5,n6){}
\drawloop[loopangle=-90.0](n6){}
\drawloop[loopdiam=7.73,loopangle=-86.01](n7){}
\node[NLangle=0.0](n8)(76.62,-8.0){$s'_{0}$}
\nodelabel[ExtNL=y,NLangle=0.0](n8){$\{a\}$}
\node[NLangle=0.0](n9)(88.36,-20.0){$s'_{2}$}
\nodelabel[ExtNL=y,NLangle=0.0](n9){$\{b\}$}
\node[NLangle=0.0](n10)(76.36,-36.0){$s'_{3}$}
\nodelabel[ExtNL=y,NLangle=0.0](n10){$\{a\}$}
\node[NLangle=0.0](n11)(68.62,-60.0){$s'_{4}$}
\nodelabel[ExtNL=y,NLangle=0.0](n11){$\{\}$}
\node[NLangle=0.0](n12)(84.62,-60.0){$s'_{5}$}
\nodelabel[ExtNL=y,NLangle=0.0](n12){$\{b\}$}
\drawedge(n8,n9){}
\drawedge(n9,n10){}
\drawedge(n10,n11){}
\drawedge(n10,n12){}
\drawloop[loopangle=-88.67](n11){}
\drawloop[loopangle=-85.82](n12){}
\node[NLangle=0.0](n40)(64.36,-20.0){$s'_{1}$}
\nodelabel[ExtNL=y,NLangle=0.0](n40){$\{\}$}
\drawedge(n8,n40){}
\drawedge(n40,n10){}
\end{picture}}
\caption{KS $\mathcal{K}$ (left) and its quotient $\mathcal{K/}_{\mathcal{R}}$ under a KME (right)}
\end{figure}
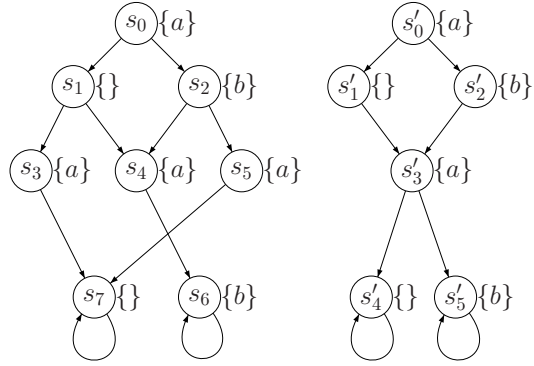
\begin{definition} Any Kripke structure $\mathcal{K}$ and its quotient $\mathcal{K/}_{\mathcal{R}}$ under KME relation $\mathcal{R}$ are $\star$-equivalent, denoted by $\mathcal{K}\star\mathcal{K/}_{\mathcal{R}}$, if and only if there exists a KME relation $\mathcal{R^{*}}$ defined on the disjoint union $S\uplus S/_{\mathcal{R}}$ such that $\forall C\in S/_{\mathcal{R}}$, $s\in C$ it holds: $(s,C)\in \mathcal{R^{*}}$.
\end{definition}
\begin{theorem}
Let $\mathcal{K}$ be a Kripke structure and $\mathcal{R}$ be a KME on $\mathcal{K}$. 
Then $\mathcal{K}\star\mathcal{K/}_{\mathcal{R}}$.
\end{theorem}
\begin{remark}
Note that KMEs are not unique, i.e., there can be more than one equivalence relation that is a KME for any given KS. Intuitively it means that the original KS $\mathcal{K}$ can be reduced in different ways.
\end{remark}
\begin{definition}[Strong bisimulation] Binary relation $\mathcal{R}$ on $S$ is a strong bisimulation on $\mathcal{K}$ if for any $(s_{1},s_{2})\in \mathcal{R}$ we have:
\begin{itemize}
 \item $L(s_{1})=L(s_{2})$,
 \item if $s'_{1}\in Post(s_{1})$ then there exists $s'_{2}\in Post(s_{2})$ with $(s'_{1},s'_{2})\in \mathcal{R}$, and
 \item if $s'_{2}\in Post(s_{2})$ then there exists $s'_{1}\in Post(s_{1})$ with $(s'_{1},s'_{2})\in \mathcal{R}$.
\end{itemize}
States $s_{1}$, $s_{2}$ are \emph{bisimilar}, denoted $s_{1}\sim s_{2}$, if $(s_{1},s_{2})\in \mathcal{R}$ for some  strong bisimulation $\mathcal{R}$.
\end{definition}
These conditions require that any two bisimilar states, say $s_{1}$, $s_{2}$ are equally labeled and that every outgoing transition of $s_{1}$ must be matched by an outgoing transition of $s_{2}$ and vice versa. Note that the relation $\sim$ is an equivalence relation and is the coarsest strong bisimulation.  
\begin{theorem}
 $\star$ is strictly coarser than $\sim$.
\end{theorem}
This theorem says that state space reduction under KME can potentially be larger than for strong bisimulation.

For strong simulation equivalence, the condition to exhibit identical stepwise behavior is slightly relaxed. Whenever $s'$ simulates $s$, state $s'$ can mimic all stepwise behavior of $s$; the reverse is not guaranteed, so state $s'$ may perform transitions that cannot be matched by state $s$. Two Kripke structures $\mathcal{K}$ and $\mathcal{K'}$ are simulation-equivalent if their initial states mutually simulate each other.     
\begin{remark}
Consider the two KSs in Fig. 2, here $\mathcal{K}$ and $\mathcal{K/}_{\mathcal{R}}$ are not strong simulation equivalent. To show that KME is strictly coarser than strong simulation equivalence, the proof of Thm. 2 can be extended showing that quotient obtained under simulation equivalence can be obtained by repeated application of KME.    
\end{remark}

\paragraph*{\textbf{Linear-time Properties.}}
We investigate linear-time properties for KSs that are preserved under KME quotienting. We study a more general class of linear-time properties that are defined over infinite words, i.e., $(2^{AP})^{\omega}$. These include, e.g., $\omega$-regular properties. Note that the preservation of $\omega$-regular properties implies the preservation of LTL formulas. These preservation results can be exploited for model checking by reducing the KS models under consideration prior to carrying out the verification.      
\begin{definition}
A \emph{linear-time} property (LT property) over the set of atomic propositions AP is a subset of $(2^{AP})^{\omega}$.
\end{definition}
\begin{example}
An LT property can be used to specify the desired behavior of the system under consideration such as:
\begin{itemize}
 \item Every time the process tries to send a message, it eventually succeeds in sending it. 
 \item Whenever the system is down, an alarm should ring until it is up again.  
\end{itemize}
\end{example}

\begin{definition}
Let $P$ be an LT property over $AP$ and $\mathcal{K}=(S,\rightarrow,AP,L,s_{0})$ a Kripke structure. Then $\mathcal{K}$ satisfies $P$, denoted $\mathcal{K}\models P$, iff $Traces^{\mathcal{K}}(s_{0})\subseteq P$. 
\end{definition}

\begin{theorem}
Let $\mathcal{K}$ be a KS and $\mathcal{R}$ be a KME on $\mathcal{K}$. Then for any LT property $P$: 
\begin{equation*}
\mathcal{K}\models P\Leftrightarrow \mathcal{K/}_{\mathcal{R}}\models P. 
\end{equation*}
\end{theorem}
Intuitively, this theorem says that if a LT property holds for the original Kripke structure, it also holds for the quotient and vice versa. In principle this result allows performing model checking on the quotient Kripke structure provided that we can obtain this in an algorithmic manner.      
\begin{corollary}
Let $\mathcal{K}$ be a KS and $\mathcal{R}$ be a KME on $\mathcal{K}$. Then for any $LTL$ formula $\varphi$:\\
\begin{equation*}
\mathcal{K}\models\varphi \Leftrightarrow \mathcal{K/}_{\mathcal{R}}\models \varphi.
\end{equation*} 
\end{corollary}
\section{Weak Kripke Minimization Equivalence}
In this section we define weak Kripke minimization equivalence (WKME). WKME is a variant of KME that abstracts from stutter steps, also referred to as internal or nonobservable steps. Note that weak equivalence relations are important for system synthesis as well as system analysis. To compare KSs that model a given system at different abstraction levels, it is often too demanding to require a statewise equivalence. Instead, a state in a KS at a high level of abstraction can be modeled by a sequence of states in the more concrete KS. Secondly, by abstracting from internal steps, quotient KSs are obtained that may be significantly smaller than the quotient under corresponding strong equivalence relation. Interestingly, though, still a rather rich set of properties is preserved under such abstractions.        
\begin{definition}[Weak predecessor based reachability]
For $s\in S$ and $C,D \subseteq S$, the function $WPbr:S\times2^{S}\times 2^{S}\rightarrow \{0,1\}$ is defined as:
\begin{equation*}
WPbr(s,C,D)=
\left\{\begin{array}{ll}
1 & \mbox{if } \exists s'\in Post(s,C), s''\in D \mbox{ s.t.}\\
  & s'\mv{*}s''\\ 
0 & \mbox{otherwise.} 
\end{array}\right.
\end{equation*} 
where $s'\mv{*}s''$ denotes that $s'$ can reach $s''$ in zero or more stutter steps, i.e., $s'\mv{}\underbrace{\ldots}_{n-times}\\\mv{}s''$ where $n\ge 0$. 
\end{definition}
\begin{definition}[WKME]
Equivalence $\mathcal{R}$ on $S$ is a \emph{weak Kripke minimization equivalence} (WKME) on $\mathcal{K}$ if we have: 
\begin{enumerate}
\item 
$\forall (s_{1},s_{2}) \in \mathcal{R}$ it holds: $L(s_{1}) = L(s_{2})$ and
\item $\forall C,D \in S/_{\mathcal{R}}$ s.t. $C\neq D$ and $\forall s', s'' \in pred(C)$ s.t. $s', s'' \notin C$ it holds:
$WPbr(s',C,D) = WPbr(s'',C,D)$.
\end{enumerate}
States $s_{1}$, $s_{2}$ are \emph{weak Kripke minimization equivalent}, denoted by $s_{1}\odot s_{2}$, if $(s_{1},s_{2})\in \mathcal{R}$ for some WKME $\mathcal{R}$. 
\end{definition}
\begin{example}
Consider the KS $\mathcal{K}$ in Fig. 3 (left). Let $C=\{s_{3},s_{4},s_{5}\}$ and $D=\{s_{6}\}$. Then $WPbr(s_{1},C,D)=1$, since it is possible to move from $s_{1}$ to $s_{6}$ in three steps via $s_{3}$, $s_{4}$ (where $s_{3}\mv{} s_{4}$ is a stutter step). Similarly $WPbr(s_{2},C,D)=1$. For KS $\mathcal{K}$, the equivalence relation induced by the partitioning $\{\{s_{0}\},\{s_{1}\},\{s_{2}\},\{s_{3},s_{4},s_{5}\},\{s_{6}\},\{\\s_{7}\}\}$ is a WKME relation.   
\end{example}
\begin{definition}[Quotient Kripke structure]
For WKME relation $\mathcal{R}$ on $\mathcal{K}$, the quotient Kripke structure $\mathcal{K/}_{\mathcal{R}}$ is defined by $\mathcal{K/}_{\mathcal{R}}=(S/_{\mathcal{R}},\rightarrow^{'},AP,L',s'_{0})$ where:
\begin{itemize}
\item 
$S/_{\mathcal{R}}$ is the set of all equivalence classes under $\mathcal{R}$,
\item $\rightarrow^{'}$ is defined by: $C\mv{}^{'}D$, s.t. $C\neq D$ iff $WPbr(s',C,D)=1$ where $s'\in pred(C)$, and $C\mv{}^{'}C$ iff there exists $s\in C$ s.t. $s\mv{+}s$
\item $L'(C)=L(s)$, where $s\in C$ and
\item $s'_{0}=C$ where $s_{0} \in C$.
\end{itemize}
where $s\mv{+}s$ denotes that $s$ can reach itself in one or more stutter steps. 
\end{definition} 
\begin{example}
The quotient KS for the Fig. 3 (left) under the WKME relation with partition $\{\{s_{0}\},\{s_{1}\},\{s_{2}\},\{s_{3},s_{4},s_{5}\},\{s_{6}\},\{s_{7}\}\}$ is shown in Fig. 3 (right).
\end{example}
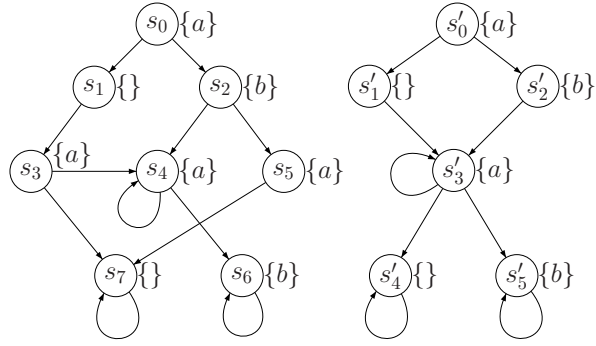
\begin{figure}
\centering
\scalebox{0.7}{
\Large\begin{picture}(111,71)(0,-71)
\node[NLangle=0.0](n0)(27.9,-4.18){$s_{0}$}
\nodelabel[NLangle=0.0](n0){}
\nodelabel[ExtNL=y,NLangle=0.0](n0){$\{a\}$}

\node[NLangle=0.0](n1)(15.9,-16.18){$s_{1}$}
\nodelabel[ExtNL=y,NLangle=0.0](n1){$\{\}$}

\node[NLangle=0.0](n2)(39.9,-16.18){$s_{2}$}
\nodelabel[ExtNL=y,NLangle=0.0](n2){$\{b\}$}

\node[NLangle=0.0](n3)(3.9,-32.18){$s_{3}$}
\nodelabel[ExtNL=y,NLangle=21.7](n3){$\{a\}$}

\node[NLangle=0.0](n4)(27.9,-32.18){$s_{4}$}
\nodelabel[ExtNL=y,NLangle=0.0](n4){$\{a\}$}

\node[NLangle=0.0](n5)(51.9,-32.18){$s_{5}$}
\nodelabel[ExtNL=y,NLangle=0.0](n5){$\{a\}$}

\node[NLangle=0.0](n6)(44.0,-52.0){$s_{6}$}
\nodelabel[ExtNL=y,NLangle=0.0](n6){$\{b\}$}

\node[NLangle=0.0](n7)(20.0,-52.0){$s_{7}$}
\nodelabel[ExtNL=y,NLangle=0.0](n7){$\{\}$}

\drawedge(n0,n1){}

\drawedge(n0,n2){}

\drawedge(n1,n3){}

\drawedge(n2,n4){}

\drawedge(n2,n5){}

\drawedge(n3,n4){}

\drawedge(n4,n6){}

\drawedge(n3,n7){}

\drawedge(n5,n7){}

\drawloop[loopdiam=7.58,loopangle=242.02](n4){}

\drawloop[loopdiam=7.53,loopangle=-88.64](n6){}

\drawloop[loopangle=-87.18](n7){}

\node[NLangle=0.0](n8)(84.88,-4.18){$s'_{0}$}
\nodelabel[ExtNL=y,NLangle=0.0](n8){$\{a\}$}

\node[NLangle=0.0](n9)(68.18,-16.0){$s'_{1}$}
\nodelabel[ExtNL=y,NLangle=0.0](n9){$\{\}$}

\node[NLangle=0.0](n10)(100.18,-16.0){$s'_{2}$}
\nodelabel[ExtNL=y,NLangle=0.0](n10){$\{b\}$}

\node[NLangle=0.0](n11)(84.18,-32.0){$s'_{3}$}
\nodelabel[ExtNL=y,NLangle=0.0](n11){$\{a\}$}

\node[NLangle=0.0](n12)(72.18,-52.0){$s'_{4}$}
\nodelabel[ExtNL=y,NLangle=0.0](n12){$\{\}$}

\node[NLangle=0.0](n13)(96.18,-52.0){$s'_{5}$}
\nodelabel[ExtNL=y,NLangle=0.0](n13){$\{b\}$}

\drawedge(n8,n9){}

\drawedge(n8,n10){}

\drawedge(n10,n11){}

\drawedge(n9,n11){}

\drawedge(n11,n12){}

\drawedge(n11,n13){}

\drawloop[loopangle=185.44](n11){}

\drawloop[loopangle=264.43](n12){}

\drawloop[loopangle=-84.14](n13){}

\end{picture}}
\caption{KS $\mathcal{K}$ (left) and its quotient $\mathcal{K/}_{\mathcal{R}}$ under a WKME (right)}
\end{figure}
\begin{definition}
Any Kripke structure $\mathcal{K}$ and its quotient $\mathcal{K/}_{\mathcal{R}}$ under WKME relation $\mathcal{R}$ are $\odot$-equivalent denoted by $\mathcal{K}\odot\mathcal{K/}_{\mathcal{R}}$ if and only if there exists a WKME relation $\mathcal{R^{*}}$ defined on disjoint union $S\uplus S/_{\mathcal{R}}$ such that $\forall C\in S/_{\mathcal{R}}$, $s\in C$ it holds: $(s,C)\in \mathcal{R^{*}}$.
\end{definition}
\begin{theorem}
 Let $\mathcal{K}$ be a Kripke structure and $\mathcal{R}$ be a WKME on $\mathcal{K}$. 
Then $\mathcal{K}\odot\mathcal{K/}_{\mathcal{R}}$.
\end{theorem}
\begin{remark}
Note that WKMEs are not unique, i.e., there can be more than one equivalence relation that is a WKME for any given KS.  
\end{remark}

\begin{theorem}
 $\odot$ is strictly coarser than $\star$. 
\end{theorem}
\begin{definition}
Let $\mathcal{K}$ be a Kripke structure and $\mathcal{R}$ an equivalence relation on $S$.
\begin{itemize}
\item $s\in S$ is $\mathcal{R}$-divergence-sensitive if there exists an infinite path fragment $\pi=s\mv{}s_{1}\\ \mv{}s_{2}...\in Paths(s)$ s.t. $(s,s_{j}\in \mathcal{R})$ for all $j>0$.
\item $\mathcal{R}$ is divergence-sensitive if for any $(s_{1},s_{2})\in \mathcal{R}$: if $s_{1}$ is $\mathcal{R}$-divergence-sensitive, then $s_{2}$ is $\mathcal{R}$-divergence-sensitive. 
\end{itemize}  
\end{definition}
\begin{definition}
Divergence-sensitive relation $\mathcal{R}$ on $S$ is a stutter bisimulation on $\mathcal{K}$ if for any $(s_{1},s_{2})\in \mathcal{R}$ we have:
\begin{itemize}
 \item $L(s_{1})=L(s_{2})$,
 \item If $s'_{1}\in Post(s_{1})$ with $(s'_{1},s_{2})\notin \mathcal{R}$, then there exists a finite path fragment $s_{2}\mv{}u_{1}\mv{}\ldots u_{n}\mv{}s'_{2}$ with $n\ge 0$ and $(s_{1},u_{i})\in \mathcal{R}$, $i=1,\ldots,n$ and $(s'_{1},s'_{2})\in \mathcal{R}$,
 \item If $s'_{2}\in Post(s_{2})$ with $(s_{1},s'_{2})\notin \mathcal{R}$, then there exists a finite path fragment $s_{1}\mv{}v_{1}\mv{}\ldots v_{n}\mv{}s'_{1}$ with $n\ge 0$ and $(v_{i},s_{2})\in \mathcal{R}$, $i=1,\ldots,n$ and $(s'_{1},s'_{2})\in \mathcal{R}$. 
\end{itemize}
States $s_{1}$ and $s_{2}$ are divergence-sensitive stutter bisimilar, denoted by $s_{1}\cong^{div}s_{2}$, if $(s_{1},s_{2})\in \mathcal{R}$ for some divergence-sensitive stutter bisimulation $\mathcal{R}$.    
\end{definition}
Next, we investigate the relationship between WKME and divergence-sensitive stutter bisimulation relation. 
\begin{theorem}
$\odot$ is strictly coarser than $\cong^{div}$. 
\end{theorem}
This theorem asserts that WKME can achieve larger state space reduction as compared to divergence-sensitive stutter bisimulation.  

For divergence-sensitive stutter simulation equivalence \cite{nejati2003} the conditions provided in Def. 19 are slightly relaxed. Whenever $s'$ stutter simulates $s$, state $s'$ can stutter mimic all stepwise behavior of $s$, and if there exists a path $\pi$ emanating from state $s$ such that all the states on $\pi$ are related to state $s'$, then $s'$ has to have some successor $s'_{n}$ such that some state $s_{n}$ on $\pi$ is related to $s'_{n}$, the reverse is not guaranteed, so state $s'$ may perform transitions that cannot be stutter mimicked by state $s$. Two Kripke structures $\mathcal{K}$ and $\mathcal{K'}$ are divergence-sensitive stutter simulation-equivalent if their initial states mutually stutter simulate each other according to the conditions given  above.     
\begin{remark}
Consider the two KSs in Fig. 3, here $\mathcal{K}$ and $\mathcal{K/}_{\mathcal{R}}$ are not divergence-sensitive stutter simulation equivalent. To show that WKME is strictly coarser than divergence-sensitive stutter simulation equivalence, the proof of Thm. 6 can be extended showing that quotient obtained under divergence-sensitive stutter simulation equivalence can be obtained by repeated application of WKME.
\end{remark}
\paragraph*{\textbf{Stutter-insensitive Linear-time Properties.}}
We investigate stutter-insensitive LT properties defined over infinite words for KSs that are preserved under WKME quotienting. These include, e.g., stutter-insensitive $\omega$-regular properties. Note that  the preservation of stutter-insensitive $\omega$-regular properties implies the preservation of $LTL/_{\bigcirc}$ formulas.      
\begin{definition}
LT property $P$ is stutter-insensitive if for any $\rho \in P$, $\forall \rho_{1}$ s.t. $\rho_{1}\triangleq \rho \Rightarrow \rho_{1}\in P$.  
\end{definition}
\begin{example}
Consider the stutter-insensitive LT property \cite{DaxKL09}:
\begin{align*}
\P_{n}:=\{w\in (2^{\{p\}})^{\omega}:&\mbox{ the number of occurrences of the sub-}\\
                                    &\mbox{word }\{p\}\emptyset \mbox{ in } w \mbox{ is divisible by $n$}\},
\end{align*}
for $n\ge 2$. Note that this property cannot be expressed using $LTL/_{\bigcirc}$.
\end{example}
The satisfaction relation for stutter-insensitive LT property $P$, i.e., $\mathcal{K}\models P$, is as in Def. 13. 
\begin{theorem}
Let $\mathcal{K}$ be a KS and $\mathcal{R}$ be a WKME on $\mathcal{K}$. Then for any stutter-insensitive LT property $P$: 
\begin{equation*}
\mathcal{K}\models P\Leftrightarrow \mathcal{K/}_{\mathcal{R}}\models P. 
\end{equation*}
\end{theorem}
\begin{corollary}
Let $\mathcal{K}$ be a KS and $\mathcal{R}$ be a WKME on $\mathcal{K}$. Then for any $LTL/_{\bigcirc}$ formula $\varphi$:\\
\begin{equation*}
\mathcal{K}\models\varphi \Leftrightarrow \mathcal{K/}_{\mathcal{R}}\models \varphi.
\end{equation*} 
\end{corollary}
\section{Synchronous Parallel Composition}
In this section we show that KME is compositional w.r.t. synchronous parallel composition (SCCS-like parallel composition \cite{CCMilner83}) of KSs. This result is useful for analyzing synchronous distributed algorithms and synchronous hardware circuits where processes progress in a lock-step fashion. For example say we want to compose a large KS $\mathcal{K}_{1}$ with another KS $\mathcal{K}_{2}$ and these KSs have $n$ and $m$ states respectively. Then the resulting KS $\mathcal{K}_{1}\otimes \mathcal{K}_{2}$ will have $m\cdot n$ states so it is worthwhile to compute this composition using a smaller KS $\mathcal{K'}$ Kripke minimization equivalent to $\mathcal{K}_{1}$. Synchronous parallel composition is also at the heart of Lustre \cite{Halbwachs}, a declarative programming language for reactive systems, and is used in many other hardware-oriented languages.   
\begin{definition}\cite{CCMilner83}
Let $\mathcal{K}_{1}=(S_{1},\rightarrow_{1},AP_{1},L_{1},s_{01})$ and $\mathcal{K}_{2}=(S_{2},\rightarrow_{2},AP_{2},L_{2},s_{02})$ be two Kripke structures. We say $s\mv{}_{i}s'$ if $(s,s')\in \rightarrow_{i}$ for $i=1,2$. The synchronous parallel composition of two Kripke structures is $\mathcal{K}_{1}\otimes \mathcal{K}_{2}=(S_{1}\times S_{2},\rightarrow,AP_{1}\cup AP_{2},L,(s_{01},s_{02}))$, where $(s_{01},s_{02})$ is the initial state, $L((s_{1},s_{2}))= L(s_{1})\cup L(s_{2})$, and $\rightarrow$ is given as follows:
\begin{equation*}
\frac{s_{1}\mv{}_{1}s'_{1}\wedge s_{2}\mv{}_{2}s'_{2}}{(s_{1},s_{2})\mv{}(s'_{1},s'_{2})}.
\end{equation*}
\end{definition}

\begin{theorem}
Let $\mathcal{K}$ be a KS and $\mathcal{R}$ be a KME on $\mathcal{K}$. Then for any Kripke structure $\mathcal{K}_{1}$:
\begin{equation*}
(\mathcal{K}\otimes \mathcal{K}_{1})\star (\mathcal{K/}_{R}\otimes \mathcal{K}_{1}).
\end{equation*}
\end{theorem}  
\section{Conclusions and Future Work}
We have presented two equivalence relations, Kripke minimization equivalence (KME) and weak Kripke minimization equivalence (WKME) on KSs. We defined the quotient system under these relations and proved that these relations are coarser than strong (bi)simulation and divergence-sensitive stutter (bi)simulation, respectively. Preservation results for LT properties and stutter-insensitive LT properties have been established under KME and WKME quotienting. Finally we show that KME is compositional w.r.t. synchronous parallel composition.

Developing and implementing an efficient quotienting algorithm is left for future work. Note that any algorithm that generates a quotient system under (weak) KME can potentially achieve a state space reduction that is larger than (stutter) (bi)simulation, but it cannot guarantee the smallest quotient system that is (stutter) trace equivalent to the original one.     
\paragraph*{Acknowledgements.}
The author would like to thank Joost-Pieter Katoen for his valuable feedback and comments. This work was supported by the European Commission under the India4EU project. 
 \begin{footnotesize}
\bibliographystyle{abbrv}
\bibliography{myBib,mybibProceedings,mybibArticles,mybibInbooks}
\end{footnotesize}
\end{document}